\documentclass[preprints,article,accept,pdftex,moreauthors]{Definitions/mdpi} 
\firstpage{1} 
\pubvolume{1}
\issuenum{1}
\articlenumber{0}
\pubyear{2022}
\copyrightyear{2022}
\datereceived{23 May 2022} 
\dateaccepted{21 June 2022} 
\datepublished{} 
\hreflink{https://doi.org/} 
\pdfoutput=1

\Title{Nuclear Modification Factor in Small System Collisions within Perturbative QCD Including Thermal Effects}

\TitleCitation{Nuclear Modification Factor in Small System Collisions within Perturbative QCD Including Thermal Effects}

\Author{Lucas Moriggi $^{1}$\orcidA{} and Magno Machado $^{2,}$*\orcidB{}}

\AuthorNames{Lucas Moriggi and Magno Machado}

\AuthorCitation{Moriggi, L.; Machado, M.}  

\address{%
$^{1}$ \quad Universidade Estadual do Centro-Oeste (UNICENTRO), Campus Cedeteg, Guarapuava 85015-430, Brazil; lucasmoriggi@unicentro.br\\
$^{2}$ \quad Instituto de Física, Universidade Federal do Rio Grande do Sul, Porto Alegre 90010-150, Brazil}

\corres{Correspondence: magnus@if.ufrgs.br}

\abstract{In this paper, dedicated to the memory of the late Prof. Jean Cleymans, the nuclear modification factors, $R_{xA}$, are investigated for pion production in small system collisions, measured by PHENIX experiment at RHIC (Relativistic Heavy Ion Collider). The theoretical framework is the transverse momentum $k_T$-factorization formalism for hard processes at small momentum fraction, $x$. 
Evidence for collective expansion and thermal effects for 
pions, produced at equilibrium, is studied based on phenomenological parametrization of blast-wave type in the 
relaxation time 
approximation.  The dependencies on the centrality and on the projectile species are discussed in terms of the behavior of Cronin peak and the suppression of  $R_{xA}$ at large transverse momentum, $p_T$. The multiplicity of produced particles, which is sensitive to the soft sector of the spectra, is also included in the present analysis.}


\begin{document}

\section{Introduction}

The transverse momentum ($p_{T}$) spectra of charged and neutral particles have been experimentally studied in proton--proton  ($pp$), proton--nucleus ($pA$), and nucleus--nucleus ($AA$) collisions from different perspectives. First of all, the~role played by this observable is highlighted as a measure of the partonic interactions in the perturbative QCD regime. One can extract from the particle spectra relevant information about the initial state dynamics of the colliding system.  On~the other hand, the~$p_T$ spectrum has been used as a probe of the collective behavior developed by the hydrodynamic expansion of the hot environment created in heavy ion collisions. A~key aspect of the problem is the emergence of thermal behavior observed even in small systems such as $pp$ and $pA$ collisions~\cite{CMS:2010ifv,ATLAS:2015hzw,CMS:2016fnw,ATLAS:2017rtr,ALICE:2016fzo,ALICE:2019avo}. The~nuclear effects are quantified by the nuclear modification factor, $R_{xA}$, which is  obtained by the ratio between the multiplicity of produced particles in the collision of a projectile $X$ off a nucleus $A$ and the scaling on the 
number of binary collisions, $N_{\mathrm{coll}}$,  $N_{\mathrm{coll}}[d^3N_{pp}/d^3p)]$, with $N_{pp}$ being the yield in proton-proton collisions.   
The~latter is expected in the case of an absence of final state nuclear medium effects. From~the experimental point of view, the~nuclear modification factor for small systems presents a suppression in the small transverse momentum region ($p_T$$\sim$1 GeV), followed by an enhancement in the intermediate momentum region  ($p_T$$\sim$2--5 GeV ), and  it finally goes to unity at large $p_T$. This~behavior is known as the Cronin effect~\cite{Cronin:1974zm}, which has been explained by different theoretical models~\cite{Wang:1998ww,Kopeliovich:2002yh,Vitev:2002pf,2distribuicoes,Hwa:2004zd,Gelis:2002nn,Albacete:2004gw,PhysRevD.68.054009,Blaizot:2004wu,Blaizot:2001nr,Jalilian-Marian:2005ccm}.

The scenario of hadron production from the decay of minijets described by 
$k_T$-factorization formalism, where $kT$ denotes the parton transverse momentum, considers that the cold matter nuclear (CNM) effects originate in the hard interaction 
 of the nuclei 
at initial states of the collision. However, the~particle production in those systems undergoes a hydrodynamics evolution to freeze-out, which modifies the corresponding  spectrum. In~\cite{Tripathy_2016,Tripathy:2017kwb,Qiao_2020}, it is argued that the $p_T$ spectrum can be described by performing a temporal separation in the relaxation time approximation (RTA) of the  Boltzmann transport equation~\cite{Florkowski:2016qig}. The~time separation corresponds to the hadrons produced in initial state hard collision and those produced in the equilibrium situation. Moreover, two-component models (thermal+hard) have been successful in describing experimental measurements~\cite{Bylinkin_2016,Giannini:2020gys,Urmossy:2015kva}. In~these approaches, the spectrum is decomposed into two parts: one related to the Boltzmann statistics and a second one characterized by the typical power-law behavior from 
perturbative quantum chromodynamics
(pQCD). Nevertheless, the~thermal nature of the spectrum has been posed in \cite{STAR:2006axp,Trainor:2007zj,Trainor:2018dfp}, where the two-component model takes into account a soft contribution coming from the longitudinal dissociation projectile-nucleus and a hard contribution due to the transverse production of jets. This seems to be enough to explain the data available without invoking collective flow. In~a previous study~\cite{Moriggi:2020qla}, this approach has been considered to describe the spectra of produced pions in lead--lead ($PbPb$) collisions at the Large Hadron Collider (LHC). It was shown that the thermal parametrization can be substantially modified by taking into account the nuclear effects present in the gluon distribution~function. 

In this paper, thermal effects are investigated in the nuclear modification factor, $R_{xA}$, for~neutral pion production, which has been measured by the PHENIX Collaboration at the Relativistic Heavy Ion Collider (RHIC) \cite{PHENIX:2021dod}, in small system collisions. A~salient feature of  $R_{xA}$ in these systems is that the Cronin peak tends to grow for smaller projectiles, which is in contradiction to the expected behavior from pQCD. In~particular, the interface between the hard process described within the QCD $k_T$-factorization formalism and the thermal sector is investigated, which can play an important role in more central collisions. To~do this, it is 
studied both the hard part of the spectrum and the multiplicity 
$dN/dy$, determined by the soft part. One~important issue regards the separation region where the collective effects are needed for 
the spectrum description. In~\cite{Rath:2019cpe}, an analysis of the high multiplicity distribution of particles in  $pp$ collisions was performed, and it was argued that the bulk part of spectra ($p_T \lesssim 2.5$ GeV) can be described by a distribution-like 
blast-wave. In~\cite{Jiang:2013gxa}, an approach based on a blast-wave parametrization that incorporates Tsallis statistics was also considered in order to study the collective flow effects in  $pp$ collisions as a function of the 
center-of-mass energy, $\sqrt{s}$. Evidence of collective expansion in high energies was shown. This suggests a dependence of the radial flow as the multiplicity or collision energy increase. Such effects also become important with the increasing number of constituents of a projectile/nuclear target. Therefore, it is fundamental to analyze the produced spectrum for distinct projectiles in order to relate the emergence of thermal behavior in terms of  $dN/dy$,$\sqrt{s}$ and the geometric parameters of the colliding system as the number of binary collisions, 
$N_{\mathrm{coll}}$, and the number of participants, $N_{\mathrm{part}}$. 

This paper is organized as follows. In~Section~\ref{sec:model1}, the predictions of pQCD for the nuclear modification factor,  $R_{xA}$, is discussed in~different approaches. The~main 
theoretical 
expressions for computing this observable are presented in the context of the pQCD $k_T$-factorization formalism. The~main physical input is the nuclear unintegrated gluon distribution (nUGD). Following~\cite{Moriggi:2020qla}, the Moriggi--Peccini--Machado (MPM) analytical parametrization for the nuclear UGD 
 is considered. It describes correctly the spectra of particles produced in $pp$ collisions, as well as the nuclear modification factors 
in $pPb$ collisions at the LHC (Large Hadron Collider). 
In~Section~\ref{sec:model2}, how the thermal corrections can be included in the spectrum is determined by using  a blast--wave model.
 The focus is on pion production in  $p+Al$, $p+Au$, $d+Au$, and $He+Au$ collisions.  It is demonstrated that the Cronin peak 
decreases 
for larger projectiles, in opposition to what is expected, as only cold nuclear matter effects are considered.  It is verified that $R_{xA}$ presents a behavior almost independent of projectile species at $p_T$$\sim$10 GeV. The~same occurs for the thermal parameters of the system, 
such as the relaxation time, $t_r$, and  temperature, $T$, which could  suggest a correlation between the energy loss at large $p_T$ and the production of a thermal system of particles.  These results and corresponding discussions are presented in Section~\ref{sec:results}. Conclusions are summarized in Section~\ref{sec:conc}.

\section{Theoretical Framework and Main~Predictions}
\unskip
\subsection{Nuclear Effects in the Gluon Distribution in a~Nucleus}
\label{sec:model1}
In the collinear factorization formalism of pQCD, the $XA$ cross-section can be expressed as the convolution of the parton distribution functions (PDFs) of the small projectile (labeled here as $X$) and the nucleus $A$ with the hard parton level cross-section. In~this context, the nuclear modification factor is described in terms of the modification of the parton distribution within a nucleus compared to the ones in a free nucleon. These modifications give rise to the  shadowing/anti-shadowing effects observed in the nuclear structure functions probed in deep-inelastic-scattering (DIS) events~\cite{Eskola,Helenius:2013bya,Helenius:2015wda,Kovarik:2015cma,NNPDFnuclear}. Moreover, the~multiple interactions among nucleons can be described in the Glauber model, where the nuclear cross-section is expected to scale with $N_{\mathrm{coll}}$. Corrections to the collinear factorization approach can take into account an intrinsic transverse momentum, $ \langle k_T \rangle$, of~partons at initial state. This effect increases 
 with $N_{\mathrm{coll}}$, which leads to an enhancement of $R_{xA}$ with respect to centrality~\cite{Wang:1998ww,Vitev:2008vk,Zhang:2001ce}.

An important aspect that appears in RHIC data~\cite{PHENIX:2021dod} is that the  Cronin 
peak tends to increase for smaller projectiles species in agreement with the ordering  
\mbox{$R_{^3He+Au} < R_{d+Au} < R_{p+Au} $.} However, the~collinear approach with cold nuclear matter effects predicts the following: $R_{^3HeAu} > R_{d+Au} > R_{p+Au}> R_{p+Al}$. This prediction is in agreement with what is observed with respect to the nuclear structure functions. There, the  shadowing/anti-shadowing contributions are intensified due to the atomic mass number,  $A$. Another essential aspect presented by these data is that for more central collisions (0--5\%), the~factor $R_{^3He+Au}$ presents a suppression that is not seen for smaller projectiles. Moreover, it is worth mentioning that  $R_{^3He+Au}$ increases for more peripheral collisions, while the factors $R_{p(d)A}$ decrease. In~the multiple scattering model, it is expected that the average transverse momentum, $\langle k_T \rangle$, acquired by partons from the projectile as a result of interaction with the nuclear target is larger in more central collisions. Therefore, this effect raises the Cronin peak, and a decrease of $R_{xA}$ is consequently anticipated in more peripheral~reactions. 

In the context of parton saturation approaches, like the color glass condensate (CGC) effective theory, the~nuclear modification factor can be associated to the saturation  of the nuclear UGD in the region of small $p_T$. This saturation is more intense in more central collisions and for large nuclei (dense color system).  In~those approaches, through the  Glauber--Mueller formalism~\cite{Glauber:1955qq,Mueller:1989st}, the multiple interactions of colored partons in the projectile with the color field of the nuclear target modifies the transverse momentum of the gluons in target, and the Cronin peak is reproduced. One can find that the~evolution in the rapidity of the nUGD and the nuclear geometry should influence the behavior of $R_{pA}$ in terms of $p_T$ \cite{2distribuicoes,PhysRevD.68.054009,Albacete:2003iq}. The~saturation approach has been used in different 
 studies~\cite{Lappi:2013zma,DHJ,Magno,BUW,Rezaeian:2012ye,Kharzeev:2004yx} in order to describe the $p_T$-spectrum and the ratio 
$R_{pA}$ 
for $dAu$ reaction at RHIC and $pPb$ ones at the LHC. In~$AA$ collisions, the saturation approach has been utilized in \cite{Armesto:2004ud,ALbacete:2010ad,KLN1,Tribedy:2011aa,Lappi:2011gu,Levin:2011hr,Albacete:2010bs,Duraes:2016yyg}.

In this study, we consider the $k_T$-factorization formalism in contrast to the collinear one. The~main advantage is that the initial partons already present non-zero transverse momenta and the effects of parton saturation can be easily implemented. Now, the~$p_T$-spectrum is given by the convolution of the unintegrated gluon distributions of the colliding nuclei, $\phi_{A,B}$, and~the production cross-section at the parton level. The~nUGD depends on the impact parameter, $\vec{b}$, of~the reaction. In~particular, the~gluon production in $A+B$ collisions at very high energies is given by the following invariant cross section:
\begin{eqnarray}
\label{eq:fatkt}
E\frac{d^3 N(b) }{dp^3}^{AB \rightarrow g+X}&=&\frac{2\alpha_s}{C_F}\frac{1}{p_T^2}\int d^2\vec{s} \, d^2\vec{k}_T\,  \phi_A(x_A,k_T^2,\vec{s}) \phi_B(x_B , (\vec{p}_T-\vec{k}_T)^2, \vec{b}-\vec{s}), 
\end{eqnarray}
where 
$x_{A,B}=(p_T/\sqrt{s})e^{\pm y}$ are the longitudinal momentum fraction of incoming partons in 
projectile $A$ and target $B$, respectively. The~quantity $\vec{s}$ is the transverse coordinate of the produced
gluon. The~strong coupling constant is $\alpha_s$, and $C_F= (N_c^2-1)/2N_c$ is the QCD Casimir color factor
with $N_c$ being the color number. Here, some remarks are in order. The processes initiated by quarks, $q_f+g \rightarrow q_f$ and $g+q_f \rightarrow q_f$,    are important in the fragmentation region and under large enough $p_T$ 
\cite{Czech:2005vp,Czech:2005vy}. Here, $g$ and $qf$ denote the gluon and fragmenting quark of flavour $f$.  The~data from PHENIX Collaboration~\cite{PHENIX:2021dod} correspond to midrapidities ($|\eta| < 0.35$) and not so large transverse momentum. This is the reason for disregarding the quark contribution in numerical~calculations here.

The unintegrated gluon distribution depends on the transverse momentum $k_T$ and longitudinal momentum fraction $x$. Here, in~the numerical calculation, we use the MPM analytical parametrization~\cite{Moriggi_2020} for the UGD. It is determined from the analysis of available data for light hadron production in $pp(\bar{p})$ collisions. It presents scaling on the variable  
$\tau=k^2_T/Q^2_s(x)$, 
based on the geometric scaling property associated to the parton saturation formalism~\cite{McLerran:2010ex,McLerran:2010wm}. The~characteristic momentum scale giving the transition between a dilute and a dense parton system is set by the saturation scale,  $Q_s$. It is defined in terms of the longitudinal momentum fraction in the form $Q_s^2(x) = (x_0/x)^{\lambda}$. The~MPM parametrization for a proton is written as follows:
\begin{eqnarray}
\label{eq:UGDp}
\phi_p(x,k_T) = \phi_p(\tau)=\frac{3\sigma_0 (1+\delta n)}{4\pi^2\alpha_s}\frac{\tau}{ \left(1+\tau \right )^{(2+\delta n)}} \ ,
\end{eqnarray}
where $\delta n = a\tau^b$, and $\lambda = 0.33$ is fixed~\cite{Praszalowicz:2015dta,Stasto:2000er}. The~parameters $\sigma_0$, $x_0$, $a$, and $b$ have been fitted from 
DESY-HERA (Deutsches Elektronen-Synchrotron, Hadron-Electron Ring Accelerator)
data for proton structure function (see discussion in \cite{Moriggi_2020}). For~central rapidity, $y$$\sim$0, the~$x$ variable can be expressed as $x=p_T/\sqrt{s}$.

The incorporation of the nuclear effects in the gluon distribution is given by the Glauber--Gribov approach for multiple scattering. 
The~main ingredients are the nuclear thickness function, $T_A(b)$, and~the color dipole cross-section. The~last quantity describes 
the multiple interactions of the leading Fock state of the projectile parton with the nucleons. The~Woods--Saxon parametrization for 
the nuclear density has been considered for a large nucleus~\cite{DeJager:1974liz,DEVRIES1987495}, and the thickness function has the 
normalization $\int d^2\vec{b} T_A(b) = A $. For~deuteron, the Hulth\'{e}n form was used~\cite{Hulthenform}, and for helium, the parametrization, presented in \cite{PhysRevC.15.1396}, is considered. Following the previous study~\cite{Moriggi:2020qla}, the~nuclear UGD is given as follows:
\begin{eqnarray}
\label{eq:fiA}
\phi_A(x,k_T^2,b)=\frac{3}{4\pi^2\alpha_s}k_T^2\nabla^2_k \mathcal{H}_0 \left\{ \frac{1-S_{q\bar{q}A}(x,r,b)}{r^2}\right\} ,
\end{eqnarray}
where $\mathcal{H}_0 \left\{ f(r) \right\}=\int rdr J_0(k_Tr)f(r)$ is the Hankel transform of order $0$.  The~quantity $\nabla^2_k$ 
is the 
two-dimensional (2-D) Laplacian in momentum space. The~key quantity is the dipole scattering matrix in configuration space, $S_{q\bar{q}A}(x,r,b)$. It can 
be determined from the cross-section for dipole scattering off a proton, $\sigma_{\rm dip}(x,r)$, in~the following way:
\begin{eqnarray}
\label{eq:SdA}
S_{q\bar{q}A}(x,r,b)&=& \exp\left[-\frac{1}{2} T_A(b)\sigma_{\rm dip}(x,r)\right], \\
\sigma_{\rm dip}^{\mathrm{MPM}}(x,r) & = & \sigma_0\left \{ 1-\frac{2\left[\left(\frac{rQ_s(x))}{2}\right)^{1+\delta n}K_{1+\delta 
n}\left(rQ_s(x)\right)\right]}{\Gamma(1+\delta n)} \right \},
\end{eqnarray}
where, in the last line, the corresponding expression for the dipole-proton cross-section in the MPM model is shown. In~particular, it scales with $rQ_s(x)$, 
as~is usual in geometric-scaling-based models. 
Here,  $K$ is the modified Bessel function of second kind and $\Gamma$ is the Gamma function. 
The~calculation is made straightforward by including any other phenomenological model for this~quantity.

Considering only CNM effects, it is found that Equation~(\ref{eq:fatkt}) predicts only small changes in $R_{xA}$ in collisions of gold nucleus $Au$ with deuteron $d$ or helium $He$. Essentially, the~nuclear modification is driven by the large nucleus in a similar way as occurs in the collinear factorization formalism. Namely, the~nuclear effects in the gluon distribution for small nuclei is tiny.  In~Section~\ref{sec:results}, the need for corrections to the initial distribution is discussed once the experimental measurements present a large difference as the projectile changes. The~nuclear modification factor,
$R_{xA}$, based on the discussion above, can be determined by using the Glauber model:
\begin{eqnarray}
\label{eq:RAA}
R_{xA}=\frac{\frac{d^3N_{XA}}{dp^3}}{N_{\mathrm{coll}} \frac{d^3 N_{pp}}{dp^3}},
\end{eqnarray}
where $d^3N_{XA}/dp^3$ is obtained by using Equations~(\ref{eq:fatkt}), (\ref{eq:fiA}), and (\ref{eq:SdA}).

The nuclear gluon distribution in Equation~(\ref{eq:fiA}) is characterized by two main quantities: the~saturation scale, 
$Q_s$$\sim$$x^{-0.33}$, which determines the increasing of the spectrum in terms of the collision energy, $\sqrt{s}$; and~a power 
index $\delta n$ that reproduces the power-like behavior on $p_T$ in the large transverse momentum limit. In Ref. \cite{Moriggi_2020}, the~investigation into the spectrum  was restricted to the region,  where 
geometric scaling is expected. For~RHIC data at 200 GeV, this is valid only at small $p_T$. In~order to achieve a good  description at large $p_T$, in~the present paper, 
the parameter $\delta n$ was modified as $\delta n =1.2$, corresponding to the effective slope observed in the spectrum in  $pp$  collisions at RHIC. On~the other hand, in~order to keep the point of maximum in the UGD at the same place, the saturation scale has to be modified as $Q^2_s(x) \rightarrow (1+\delta n) Q^2_s(x)$.  The~fragmentation process is described in a simplified way by taking the hadron transverse momentum in the approximate form  $p_{Th} = \langle z \rangle p_{Tg}$. Here, $p_{Tg}$ is the transverse momentum of the produced gluon, with~$\langle z \rangle=0.75$. The~spectrum determination contains uncertainties, such as the appropriated description of the fragmentation process, the~contribution of processes initiated by quarks at large $p_T$, and the determination of the mass of the gluon jet. However, in~the ratio   $R_{xA}$, these uncertainties are reasonably canceled, and the nuclear modification factor essentially provides the ratio of the UGDs in the nucleus and in the nucleon, $\phi_A(x,p_T,b)/ \phi_p (x,p_T)$, at a given~centrality. 

Having determined the basic parameters and the formalism to compute the unintegrated gluon distribution in both nucleons and nuclei in the initial hard process, in the next Section, we present how the  thermal effects are incorporated in the~analysis.

\subsection{Collective Expansion and the Blast-Wave~Model}
\label{sec:model2}

In \cite{Tripathy_2016,Tripathy:2017kwb,Qiao_2020}, it has been argued that the  $p_T$ spectrum can be described by doing a time separation in the RTA 
of the Boltzmann transport equation~\cite{Florkowski:2016qig}. In~order to include the thermal corrections to the 
spectrum, predicted 
by Equation \eqref{eq:fatkt}, We assume that the final state distribution $f_{\mathrm{fin}}$ can be expressed in the RTA approximation,
\begin{equation}
\label{eq:ffin}
f_{\mathrm{fin}}=f_{\mathrm{eq}}+(f_{\mathrm{in}}-f_{\mathrm{eq}})e^{-t_f/t_r},  
\end{equation}
where $t_f/t_r$ is the ratio of the freeze-out and 
relaxation times, with~$f_{\mathrm{in}}$ given by Equation~(\ref{eq:fatkt}) and $f_{\mathrm{eq}}$ being the equilibrium distribution of the thermal system. 

One way of investigating the collective properties on the spectrum is based in phenomenological models of blast-wave type~\cite{Schnedermann:1993ws}, developed in order to capture the essential aspects of the thermal and hydrodynamic description of  $AA$ collisions. These models have been applied to heavy-ion collisions at  RHIC~\cite{Abelev:2008ab,STAR:2003jwm,STAR:2017sal}, LHC~\cite{Abelev:2013vea,ALICE:2019hno,ALICE:2013wgn,ALICE:2018pal} and 
CERN/SPS (European Organizationfor Nuclear Research/Super Proton Synchrotron)~\cite{Bearden:1996dd}. 
 The~velocity profile of the expanding medium is parametrized in the following way:
\begin{equation}
\label{eq:rho}
\rho=\tanh^{-1}(\beta_T) =\tanh^{-1}\left[ \left( \frac{r}{R}\right)^m \beta_s \right],
\end{equation}
where $\beta_T$ is the transverse expansion velocity, $m$ is the velocity profile’s exponent, and $\beta_s$ is the transverse expansion velocity at the surface. The~ average speed is $\left< \beta \right>=\frac{2}{2+m}\beta_s$, $r$ is the radial distance in the transverse plane from the centre of the fireball, and $R$ is the fireballs radius. 
 Here, a linear profile is considered, that is,~$m=1$. The~spectrum of produced particles is given by the following:
\begin{equation}
\label{eq:BGBW}
f_{\mathrm{eq}}\propto m_T\int_0^R rdrK_1\left( \frac{m_T\cosh (\rho))}{T}  \right )I_0\left( \frac{p_T\sinh (\rho))}{T}  \right ),
\end{equation}
where $I_0$ ard $K_1$ are the  modified Bessel functions of the first and second kind, respectively, and~ $T$ is the kinetic freeze-out temperature, $T_{\mathrm{kin}}$, in~the context of the  Boltzmann-Gibbs blast-wave (BGBW) approach. In~the RTA approximation, considered here, $T=T_{\mathrm{eq}}$ is the temperature characterizing the Boltzmann local equilibrium distribution, $f_{\mathrm{eq}}$. It has been shown in \cite{Qiao_2020} that for a given centrality, the~temperature $T_{\mathrm{eq}}$ is larger than the kinetic one. This  is consistent with the idea that the temperature decreases with the evolution of the system from the local equilibrium to the kinetic freeze-out. The~spectrum is determined by the parameters $\langle\beta\rangle$ and $T$, which are adjusted from the experimental~measurements.

As an exploratory study, the~fitted parameters are presented in Table~\ref{tab:pars}. A~further study on possible constraints for the parameter ranges and discussion of data statistics is deserved. The~expansion average velocity varies very little for the different systems. One obtains  $\langle \beta\rangle$$\sim$0.55, which is closer to the values determined in paper \cite{Moriggi:2020qla} for pions in $PbPb$ collisions in $\sqrt{s}=2.76$ TeV  at the LHC. 
It is also noticed that in \cite{Rath:2019cpe}, the obtained values are $\langle \beta\rangle$$\sim$0.65 in $pp$ collisions, independently of the multiplicity. The~temperature follows a trend similar to that observed for  $R_{xA}$ at large $p_T$ (i.e.,~almost independent of the projectile species). This could suggest a relation between the energy loss in the large transverse momentum region and the equilibrium temperature of the~system.

\begin{table}[H]
\caption{Adjusted parameters' equilibrium temperature, which characterizes the Boltzmann local 
equilibrium distribution, $T=T_{\mathrm{eq}}$;  the~average speed, 
$\langle \beta\rangle$; and the ratio of the freeze-out and relaxation times, $t_f/t_r$. 
Results are presented for three classes of~centrality. \label{tab:pars}}

		\newcolumntype{C}{>{\centering\arraybackslash}X}
		\begin{tabularx}{\textwidth}{c|ccc|cccc|}
\hline
\multirow{2}{*}{}      & \multicolumn{3}{c|}{(0--5)\%}                                        & \multicolumn{4}{c|}{(0--20)\%}                                                                  \\ \hline

                       & \multicolumn{1}{l|}{$p+Al$} & \multicolumn{1}{l|}{$p+Au$} & $He+Au$ & \multicolumn{1}{l|}{$p+Al$} & \multicolumn{1}{l|}{$p+Au$} & \multicolumn{1}{l|}{$d+Au$} & $He+Au$  \\ 
$T$ (GeV)              & \multicolumn{1}{l|}{0.054}  & \multicolumn{1}{l|}{0.055}  & 0.041   & \multicolumn{1}{l|}{0.043}  & \multicolumn{1}{l|}{0.046}  & \multicolumn{1}{l|}{0.045}  & 0.035      \\ 
$\langle \beta\rangle$ & \multicolumn{1}{l|}{0.579}  & \multicolumn{1}{l|}{0.587}  & 0.620   & \multicolumn{1}{l|}{0.558}  & \multicolumn{1}{l|}{0.588}  & \multicolumn{1}{l|}{0.601}  & 0.608      \\ 
$t_f/t_r$              & \multicolumn{1}{l|}{0.223}  & \multicolumn{1}{l|}{0.301}  & 0.528   & \multicolumn{1}{l|}{0.223}  & \multicolumn{1}{l|}{0.223}  & \multicolumn{1}{l|}{0.288}  & 0.357    \\ 
\hline
\end{tabularx}
\end{table}
\unskip

\begin{table}[H]
		\newcolumntype{C}{>{\centering\arraybackslash}X}
		\begin{tabularx}{\textwidth}{c|cccc|}
\hline
\multirow{2}{*}{}      &\multicolumn{4}{c|}{(20--40)\%}                                                                    \\ \hline
                       &  \multicolumn{1}{l|}{$p+Al$} & \multicolumn{1}{l|}{$p+Au$} & \multicolumn{1}{l|}{$d+Au$} & $He+Au$ \\ 
$T$ (GeV)              & \multicolumn{1}{l|}{0.032}  & \multicolumn{1}{l|}{0.028}  & \multicolumn{1}{l|}{0.045}  & 0.034   \\ 
$\langle \beta\rangle$ &  \multicolumn{1}{l|}{0.457}  & \multicolumn{1}{l|}{0.508}  & \multicolumn{1}{l|}{0.557}  & 0.473   \\ 
$t_f/t_r$              & \multicolumn{1}{l|}{0.105}  & \multicolumn{1}{l|}{0.105}  & \multicolumn{1}{l|}{0.105}  & 0.105   \\ 
\hline
\end{tabularx}
\end{table}
\unskip

\section{Results and~Discussions}
\label{sec:results}

\begin{figure}[H]
\includegraphics[width=\linewidth]{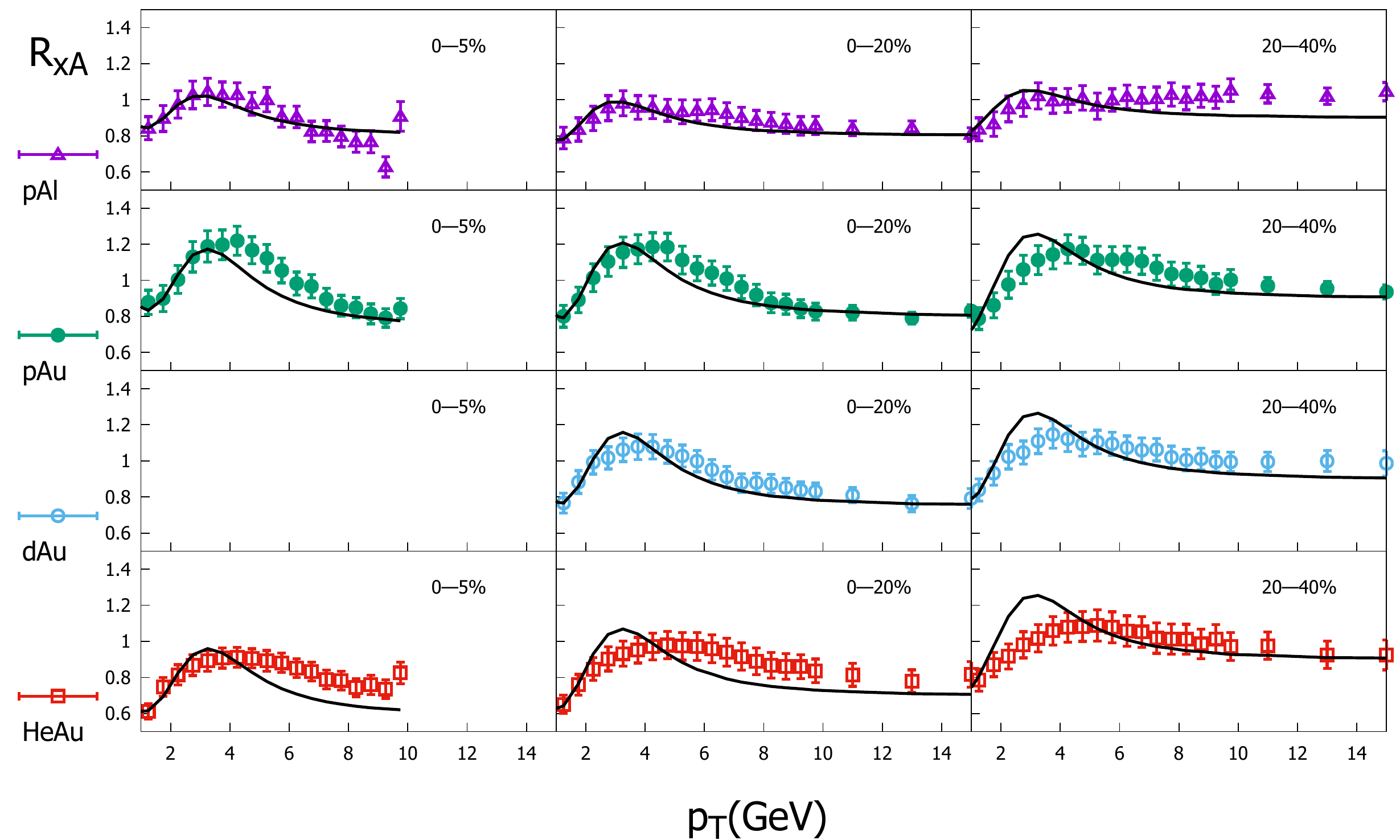}
\caption{The nuclear modification factor, $R_{xA} (p_T)$, computed using Equations~(\ref{eq:RAA})--(\ref{eq:BGBW}), with 
inclusion of thermal effects and comparison to data from PHENIX Collaboration \cite{PHENIX:2021dod}. Different projectile species and three more central  classes of centrality are~presented.} \vspace{-10pt}
\label{fig:RxA} 
\end{figure}

\begin{figure}[H]

\begin{adjustwidth}{-\extralength}{0cm}
\centering 
\includegraphics[width=1\linewidth]{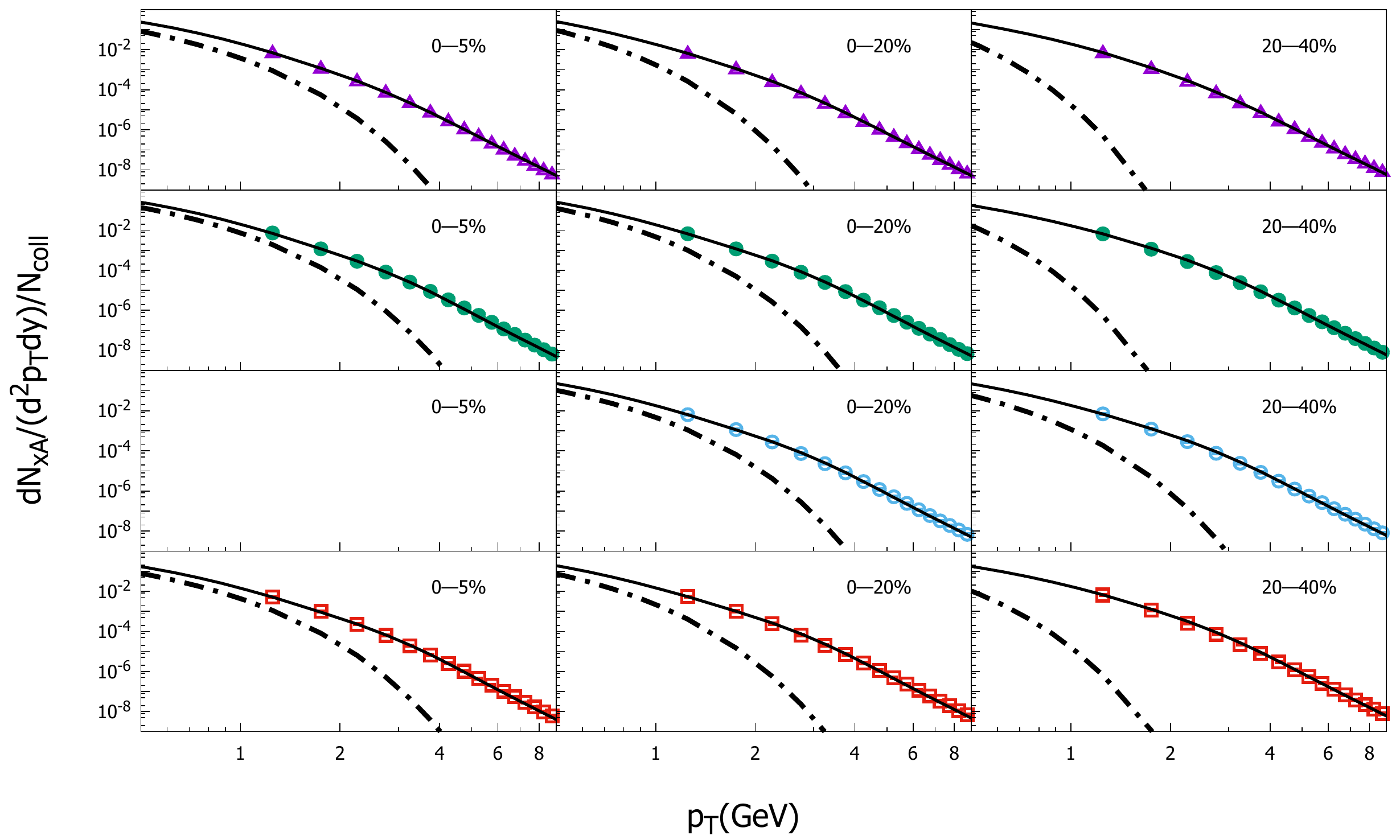}
\end{adjustwidth}
\caption{The yield  $dN_{xA}/d^2p_Tdy$ for $xA$ collisions divided by the number of binary collisions, $N_{\mathrm{coll}}$, as~a 
function of transverse momentum. Data from PHENIX Collaboration~\cite{PHENIX:2021dod} at small transverse momentum region. Different projectile species and three more central  classes of centrality are presented. The~dot-dashed lines represent the contribution of the particle produced in thermal equilibrium, and the solid lines correspond to the sum given by  Equation~(\ref{eq:ffin}). See text for details.} 
\label{fig:dndy} 
\end{figure}

Our analysis is restricted to more central collisions (0--5, 0--20, 20--40)\%, where the nuclear effects are more prominent. The~average values of geometric parameters such as $\langle N_{\mathrm{coll}} \rangle $ and $\langle N_{\mathrm{part}} \rangle $ are calculated using a Glauber MC simulation, and they are explicitly presented in the PHENIX Collaboration papers (see Table II of \cite{PHENIX:2021dod} and page 6 of \cite{PHENIX:2018hho}). In~Figure~\ref{fig:RxA}, the nuclear modification factor  $R_{xA}$ is shown for different systems, namely $p+Al$, $p+Au$, $d+Au$, and $He+Au$ . For~semi-central collisions such as (20--40)\%, $R_{xA}$ is close to unity at large $p_T$. This indicates the absence of relevant nuclear effects, which is independent of the projectile species. On~the other hand, for~the most central collisions, (0--5)\%, there exists a suppression of $R_{xA}$ at large  $p_T$ for all projectiles. This implies that the underlying mechanism of suppression does not depend on the colliding nucleus and has a strong dependence on centrality. In~the figure, the~lines represent the result obtained from  Equations~(\ref{eq:RAA})--(\ref{eq:BGBW}), including the thermal effects. In~the relaxation time approximation, $R_{xA} $$\sim$$e^{-t_f/t_r}$, the initial distribution scales with $N_{\mathrm{coll}}$. This indicates that the relaxation time diminishes for more central collisions for any projectile. This can be understood supposing that the time should be inversely proportional to the energy density, which varies little for different colliding systems. For~more peripheral reactions, the predicted Cronin peak is larger than that experimentally observed in the  $xAu$ collisions, which is related to the impact parameter dependence of the nuclear UGD. It decreases more slowly with the impact parameter than in case of the aluminium nucleus. In~the small $p_T$ region, it is observed a suppression due to the nuclear shadowing in the nUGD, followed by an enhancement near the  maximum of the nUGD.  In~the initial distribution, an enhancement in the peak and a stronger shadowing effect are expected for the $He$ nucleus. However, it was verified that the thermal contribution at small  $p_T$ contributes to diminishing these effects. A similar phenomenon occurs in heavy-ion collisions, as the produced particles in thermal equilibrium compensate the expected suppression due to strong nuclear shadowing from the nuclear UGD.  In~the case of small systems that are investigated here, the contribution of this piece, $f_{\mathrm{eq}}$, to~the spectrum is found to be small, $\sim 10\% $. This contribution is made clear if one looks at Figure~\ref{fig:dndy}, where the small $p_T$ region of the spectrum is shown.  There, the~yield  $dN_{xA}/d^2p_Tdy$ has been divided by $N_{\mathrm{coll}}$ and presented as a function of the transverse momentum. The~dot-dashed lines represent the contribution of the particle produced in thermal equilibrium, and the solid line corresponds to the sum given by  Equation~(\ref{eq:ffin}). For~less central collisions, where  $R_{xA}$ tends to unity at large $p_T$, such a contribution is small, and it is higher in more central collisions.

\begin{figure}[H]
\includegraphics[width=0.95\linewidth]{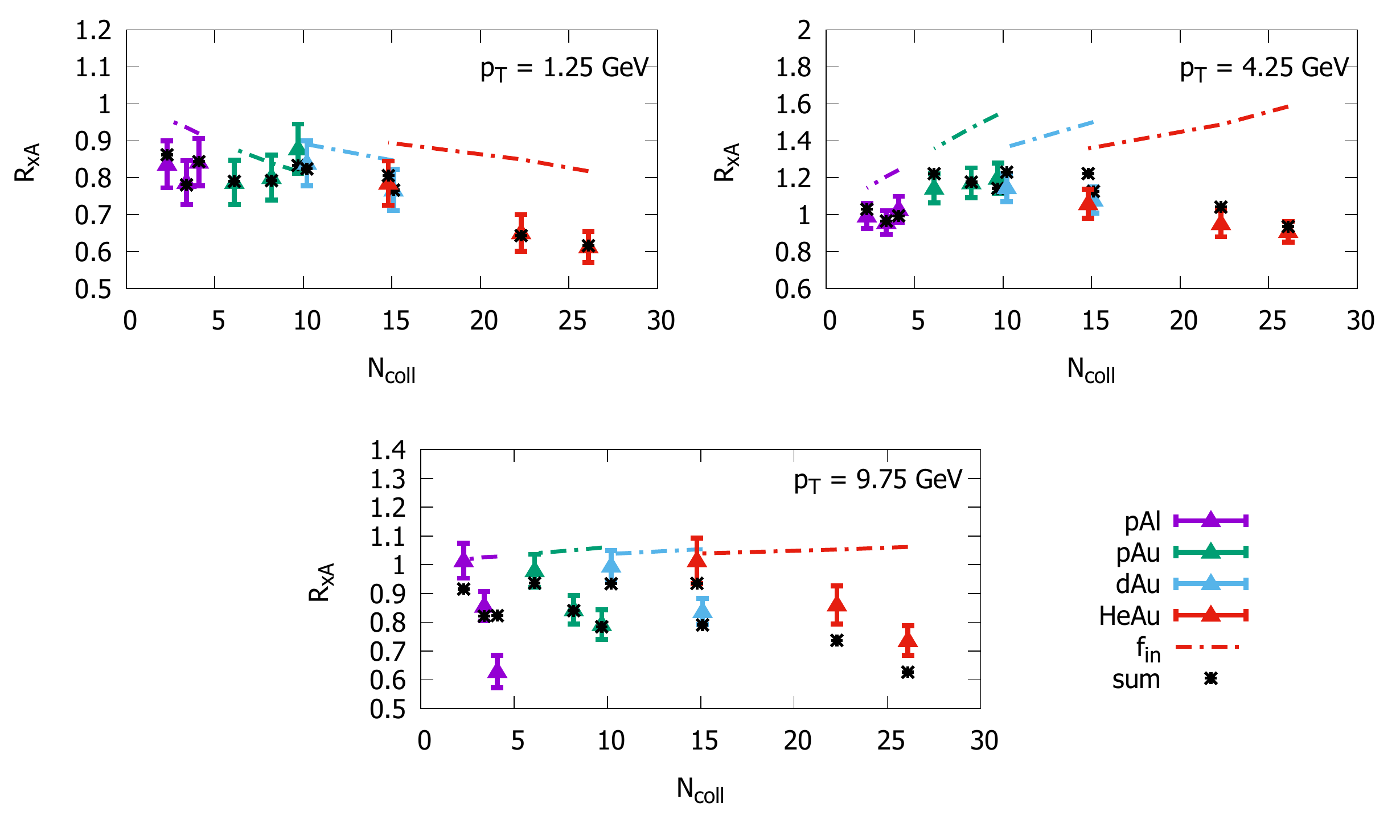}
\caption{Nuclear modification factor as a function of the number of binary collisions for three values of transverse momentum, $p_T$. Dot-dashed lines represent the calculation without thermal effects by using only $f_{\mathrm{in}}$. The~black ``$\times$'' symbols correspond to the full calculation given by Equations~(\ref{eq:RAA})--(\ref{eq:BGBW}). Data from PHENIX Collaboration~\cite{PHENIX:2021dod}.}
\label{fig:RPT}  \vspace{-22pt}
\end{figure}

Each part of the spectrum has a distinct behavior in terms of the geometric parameter of the nuclear collision. The $R_{xA}$ is analyzed in three different regions as a function of  $N_{\mathrm{coll}}$, as presented in Figure~\ref{fig:RPT}. At~$p_T=1.25$ GeV, where $R_{xA}<1$, the dot-dashed lines correspond to the expected result by considering only $f_{\mathrm{in}}$, which is determined by the nuclear shadowing present in the nuclear UGD.  We draw attention to the fact that the contribution to $R_{xA}$ from $f_{\mathrm{in}}$ presents fewer theoretical uncertainties compared to the absolute spectrum, as they are canceled in the ratio. In~this limit, $R_{xA}$ basically resembles the behavior of the ratio $\phi_A/\phi_p$ at a given centrality class.  For~$N_{\mathrm{coll}}\gtrsim 10$, data present the expected behavior, with a stronger suppression for increasing $N_{\mathrm{coll}}$. However, in~the case of  $N_{\mathrm{coll}}\lesssim 10$, the ratio  $R_{xA}$ is almost flat. For~instance, in~the case of $p+Al$, reaction  $R_{xA}$ increases for more central collisions compared to the peripheral ones. At~$p_T=4.25$ GeV, near~the Cronin peak, the~expected behavior is peak enhancement as  $N_{\mathrm{coll}}$ increases. However, for~ $N_{\mathrm{coll}}\gtrsim 10 $, the observed behavior is the opposite; namely, the~peak diminishes as the number of collisions increases. An~interesting case is the nuclear modification factor at large $p_T$. The value  $p_T=9.75$ GeV is considered, where  $R_{xA}$$\sim$$1$ is expected due to the $N_{\mathrm{coll}}$-scaling predicted by pQCD. However, the~nuclear modification factor is substantially smaller than unity for the two more central classes of centrality. The~most intriguing fact is that  $R_{xA}$ is independent of the projectile species and practically the same for each centrality. This suggests that  $R_{xA}$ in the large $p_T$ region is testing observables, which have a poor dependence on the projectile or collision geometry, such as the multiplicity or transverse energy density. In~\cite{Gu:2022xjn}, it was demonstrated that the obtained temperature in the  Tsallis-blast-wave model scales with $dN/\eta$ for  $pp$, $pA$, and $AA$ collisions but presents an intense reliance on the size of the colliding~system.

\begin{figure}[H]
\includegraphics[width=0.95\linewidth]{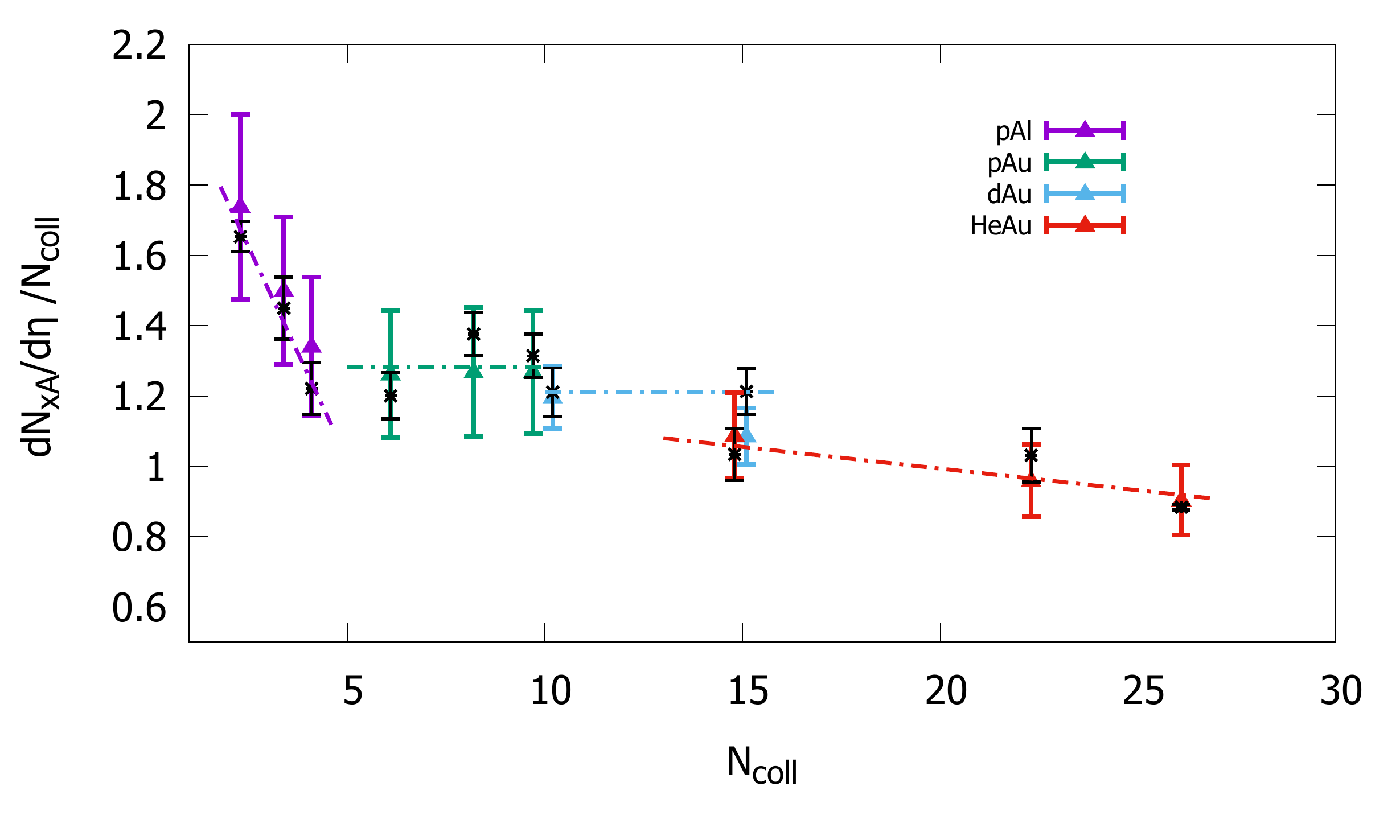} \vspace{-6pt}
\caption{Multiplicity of charged hadrons measured by PHENIX Collaboration~\cite{PHENIX:2018hho} as a function of $N_{\mathrm{coll}}$ for different projectile species and centrality classes. The black ``$\times$'' symbols represent the numerical
results, and for better viewing, the dot-dashed lines correspond to their linear regression for each projectile~species.} \vspace{-22pt}
\label{fig:TBGBW} 
\end{figure}

Finally, the quantity $dN_{\rm ch}/d\eta$ is studied, assuming that the charged hadron multiplicity $N_{\rm ch}$) is given predominantly by pions. 
Accordingly, the~integrated pion spectra (after transformation $y\rightarrow \eta$) can be compared to the measured multiplicities. 
A~comparison between predictions and the PHENIX data~\cite{PHENIX:2018hho} is presented in Figure~\ref{fig:TBGBW}. In 
the~figures, the~black ``$\times$''symbols represent the   
results, and for better viewing, the dot-dashed lines correspond to their linear 
regression for each projectile species.   For~$N_{\mathrm{\rm coll}}\gtrsim 10$, the ratio $dN_{\rm ch}/d\eta/N_{\mathrm{coll}}$ 
follows 
the same pattern observed for $R_{xA}$ at small $p_T$; namely, there is a scaling that is less steep on  $N_{\mathrm{coll}}$. In~the 
intermediate region, $dN_{\rm ch}/d\eta/N_{coll}$ is practically constant in terms of $N_{coll}$. In~\cite{PHENIX:2015vqa}, an 
analysis 
of the energy loss parameter, $\delta p_T$, for~different systems in $AA$ collisions at  RHIC and LHC (200 GeV) demonstrates that the 
energy loss imposed by the nuclear medium does not scale with the geometric parameters of the system ($N_{\mathrm{coll}}$,\, 
$N_{\mathrm{part}}$), but scales with quantities related to the system energy density and the multiplicity $dN_{\rm ch}/d\eta$.

\section{Summary and~Conclusions}
\label{sec:conc}

In this paper,the relevance of the thermal effects is investigated in collisions of small systems based on the analysis of the transverse momentum spectra of neutral pions measured at RHIC (Relativistice Heavy Ion Collider. The~ Boltzmann equation in the relaxation time approximation has been considered.  It was shown that the hard part contribution attributed to an initial production computed within the $k_T$-factorization formalism in 
pQCD (perturbative quantum chromodynamics) presents a different behavior from the one experimentally observed, even at large 
transverse momentum, $p_T$. The~deviation can be understood in terms of enhancement/suppression of particles produced in the collective expansion of the thermal system. In~particular, the~Cronin peak tends to decrease in the case of larger-size projectiles, in opposition to what is expected, due to considering cold nuclear matter effects only. It is verified that the nuclear modification factor, $R_{xA}$,
 at $p_T$$\sim$$10$ GeV does not  depend on the projectile species . The~same occurs for the thermal parameters of the system, $t_{r,T}$, which suggests a correlation between the energy loss at large $p_T$ and the production of a thermal system of~particles.



\vspace{6pt} 



\authorcontributions{L.M. and M.M. have contributed to the study equally, starting from the conceptualization of the problem, to the methodology, paper writing, and review. All authors have read and agreed to the published version of the~manuscript.}

\funding{This research was funded by the Brazilian National Council for Scientific and Technological Development (CNPq) under the contract number 306101/2018-1.}

\dataavailability{The data used can be found in the corresponding references.} 


\conflictsofinterest{The authors declare no conflict of~interest.}

\begin{adjustwidth}{-\extralength}{0cm}

\reftitle{References}



\end{adjustwidth}
\end{document}